\documentclass[12pt]{iopart}
\usepackage{graphicx}
\begin{document}
\title[Treatment time vs. plan quality in rotational radiation delivery]{On the tradeoff between treatment time and plan quality in rotational arc radiation delivery}
\author{David Craft and Thomas Bortfeld}
\address{Department of Radiation Oncology, Massachusetts General Hospital and 
Harvard Medical School, Boston, MA 02114, USA}
\ead{dcraft@partners.org}
\date{March 5, 2009}
\begin{abstract}
A new delivery option for cancer centers equipped with linear accelerators fitted with multi-leaf
collimators (MLC) -- i.e. centers which can perform intensity modulated radiation therapy (IMRT) -- is rotational
delivery.  In rotational delivery, the beam is 
on while the gantry is rotating and the MLC leaves are moving, thus treating the patient more efficiently
(regarding time) than in IMRT.  A consideration
that should be examined when evaluating this new type of delivery method is 
the tradeoff between treatment plan quality and delivery time:  what do we 
sacrifice in terms of plan quality by guaranteeing a 2 minute delivery, for example?  In this paper we
examine this question by studying a simplified 2D phantom where leaf and gantry motion are directly included 
in the optimization model.  We formulate the model as a linear mixed integer program.  Because of the difficulty
in solving to optimality, we employ additional models which allow us to trap the true optimal solution
between upper and lower bounds.  The lower bound solutions reveal which beam directions are most useful for 
treatment (i.e. where the gantry should slow down), and this information is used to produce deliverable solutions 
close to the lower bound solutions.  For the phantom cases studied, we show that when time is not an issue, IMRT
solutions are optimal, but when allowable treatment time is constrained sufficiently, 
rotational delivery is the preferred choice.
\end{abstract}

\section{Introduction}
A current trend in radiation delivery is rotational therapy, where the gantry
is continuously revolved around the patient while the beam is on.  
Depending on the specifics of the delivery (e.g. full 360 degree rotation, partial angle 
sweep, or multiple arcs), rotational therapy is variously called volumetric modulated arc therapy (VMAT),
intensity modulated arc therapy (IMAT) \cite{imat}, arc-modulated radiation therapy (AMRT) \cite{amrt}, 
or just arc therapy.  Hardware vendors pushing this technology, such
as Elekta and Varian, claim that rotational therapy is faster and the plans are as good as or better than more 
traditional delivery techniques such as 3D conformal or intensity modulated radiation therapy (IMRT), and many
presentations and papers are coming out which compare rotational plans to 3D and IMRT plans 
from a dosimetric perspective (see e.g. \cite{palma}).

The same machines that deliver IMRT and 3D conformal can deliver rotational arc therapy.  Thus, a question
that radiation therapy centers face is: which type of delivery shall we use?  In this paper,
we approach this question from the perspective of trading off treatment time versus plan quality.  If one is not
concerned with treatment time, allowing full intensity modulation at each angle offers the most flexibility
and thus is the desired method (if a particular angle is not helpful, a well designed optimizer
will detect this and not use that angle).  On the other hand, if treatment time is of concern, then it is not clear
if one should do intensity modulation from a few selected angles (i.e. IMRT) or one should use rotational 
delivery \cite{bortfeldwebb}.  (In practice there is a limit to how much a rotating gantry can slow down, and thus
there is a distinction between IMRT and rotational delivery.  In the model we present the distinction is somewhat blurred
as we allow the gantry to slow down arbitrarily, thus rotational delivery plans can have IMRT-like characteristics.  
Nevertheless, the solutions we find generally show a leaning towards
either IMRT or arc-therapy.)  Our modeling approach is to have the delivery style that gets used, IMRT or rotational delivery,
be an output of the optimization rather than an input.
In words, our model solves the problem of: given a certain treatment time limit and a set of dose 
constraints and a dose-based objective function, use the radiation delivery hardware in any way within its
capabilities to minimize the objective function.  


Solving this problem for clinical problem instances is at this point not possible due to the computational burden.
Thus we view this work as a step towards understanding and optimizing this complex issue, 
and apply it to 2D phantom geometries.  Even for these simplified models, we can solve the optimizations only 
approximately, by solving related optimization problems 
that give tight bounds.  The near-optimal solutions produce
insight into the choice between rotational delivery and IMRT.  These solution techniques
will ideally be useful in designing new clinical algorithms for optimizing rotational delivery plans, and we 
see this present work as a step
in the direction of considering treatment time directly in radiotherapy plan optimization, similar to 
the inclusion of total monitor units directly into IMRT planning \cite{craft-spg}.

\section{Optimization models}
We run two main optimization models to understand the tradeoff between treatment time $T$ and plan quality.
We would like to use only one, an exact model that includes $T$ as a constraint and lets the machine
do whatever it can to minimize the dosimetric objective function.  It turns out that this model,
even for the simplified patient and gantry/MLC model that we use, is too difficult to solve, and so
we restrict it in some simple ways, described below.  To verify that
our restrictions are not overly prohibitive, we form a second model which provides a lower bound to
how low the true objective function could go for a fixed $T$.  Together, these models allow us to pin the optimal
solution between upper and lower bounds.

The dosimetric optimization model that we use for all models is:
\begin{eqnarray}
\mathrm{ minimize~} & f(d) \nonumber \\
\mathrm{s.t.}~~&  d= Dx \nonumber \\
~ & d \in C_d \nonumber \\
~ & x \ge 0.
\label{basemod}
\end{eqnarray}
$D$ is the precomputed dose-influence matrix that relates the beamlet intensities $x$
to the resulting voxel doses $d$.  $f(d)$ is a convex function on the dose vector
and $C_d$ is a convex set for constraining the dose vector (the exact formulations are given in the Results section).
In this model $x$ is only constrained to be above 0, allowing full fluence modulation at each angle.

To make this model useful for VMAT (from here onward, we will use VMAT to mean rotational delivery while the MLC leaves
are moving), we next include mechanical delivery considerations.  
We discretize time, gantry positions, and leaf 
positions.  Using values from \cite{otto}, we assume that the maximum leaf speed 
is 1/2 centimeter per degree of gantry rotation.  Letting $\Delta$ be the gantry angle discretization size, i.e. 
$\Delta$ is maximum angle sweep (in degrees) the gantry can make in one time unit, we have that the maximum leaf motion 
over that sweep is $s = 1/2 \Delta$ cm.  At each time step, each leaf can stay still or move one
step (size $s$) in either direction (subject to collision avoidance and maximal leaf position constraints).
Similarly, at each time step, the gantry can stay still or move one step in either direction.  Thus, slowing down 
of the gantry is modeled by the gantry remaining in one angular position for more than one time step.

We consider a 2D slice of patient geometry, and we assume the MLC leaves are in line with the gantry motion.
Let $M$ equal the number of beamlets (beamlets are of size $s$) across the gantry head.  Let $T$ equal the total
number of time units allowed for treatment, and let $A$ be the number of angular positions modeled.

\subsection{Both-ways exact MIP model}
In the both-ways model, we allow the gantry to stay still, move forward, or reverse direction at each time step.
We restrict the leaves to move at most one beamlet position at each step.  We introduce the following
binary variables: $b[i,j,a,t]$.  The indices $i$ and $j$ represent the discrete index position of the left and right
leaf edges respectively, from 0 up to $M$.  Note that collision avoidance is modeled by insisting that
$i \le j$, and if $i=j$ then the leaf is closed.
The index $a$ is the angular position index, and $t$ is the time index.
If $b[i,j,a,t]=1$, this means that at time $t$, the gantry is at angular index $a$, the edge of left leaf
is at position $i$, and the edge of the right leaf is at position $j$.

With this as our underlying decision variable, we can now impose delivery constraints.  The simplest constraint to impose
is that there is only one leaf position/angular location delivered at each time, and this is handled by including the following
constraint:
\begin{equation}
\sum_{i=0}^M \sum_{j=i}^M \sum_{a=1}^A b[i,j,a,t]= 1,~~~~~~ t=1\ldots T. 
\label{oneatatime}
\end{equation}

Another constraint to impose is that the next angle and leaf position must be attainable within the constraints of the hardware.
Due to the way the discretization has been done, this is fairly straightforward.  Assume we are at time $t$. The following 
constraint forces all $b[l,r,a,t]$ to be zero except the ones that are attainable from the previous position at $t-1$:
\begin{eqnarray}
 b[l,r,a,t] \le \sum_{i= \mathrm {max}(0,l-1)}^{l+1} \sum_{j = \mathrm {max}(i,r-1)}^{\mathrm {min}(M,r+1)} b[i,j,a,t-1] + ~~~~~~~~~~~~~~~~~~~~~~~\nonumber \\
~~~~~~~ \sum_{i = \mathrm {max}(0,l-1)}^{l+1} \sum_{j = \mathrm {max}(i,r-1)}^{\mathrm {min}(M,r+1)} b[i,j,\mathrm {max}(1,a-1),t-1]+ ~~~~~~~~~~~~~~\nonumber \\
~~~~~~~~~~~~~~~~~~~~~\sum_{i = \mathrm {max}(0,l-1)}^{l+1} \sum_{j = \mathrm {max}(i,r-1)}^{\mathrm {min}(M,r+1)} b[i,j,\mathrm {min}(a+1,A),t-1]
\label{validNextAp}
\end{eqnarray}
This constraint is imposed for all $l=0\ldots M$,  $r=l\ldots M$,  $a=1\ldots A$, and $t=2\ldots T$.  The $i$ and $j$
sums are over all leaf positions at most one step in either direction, 
and accounting for the fact that the leaves have to stay within the bounds.

Finally, it is necessary to relate binary variables $b$ to the beamlet vector $x$ in the dosimetric 
optimization model (\ref{basemod}).  This is done by: 
\begin{equation}
x[k] = \sum_{t = 1}^T \sum_{i = 0}^{\mathrm {mod}(k-1,~M)} \sum_{j = \mathrm {mod}(k-1,~ M+1)}^M b[i,j,\mathrm {floor}((k-1)/M)+1,t],~k = 1 \ldots M\cdot A 
\label{setx}
\end{equation}
Here the $i$ and $j$ sums yield all the leaf positions for which the particular beamlet is exposed.

This model can easily be modified to, for example, restrict the gantry motion to one direction, either with or without the option
to slow down.  Restricting the gantry to maintain a steady speed simplifies the model enough so that it is computationally feasible for
our phantom geometry.  This is done by coupling the time and angle indices (i.e. $a=t$).  But this is fairly restrictive and so
instead we use the following model
which is less restrictive than $a=t$, but not as difficult to solve as the above both-ways model.

\subsection{Possible-previous-angle exact MIP model (PPA)}
\begin{figure}[ht] 
\centerline{\includegraphics[width=10cm]{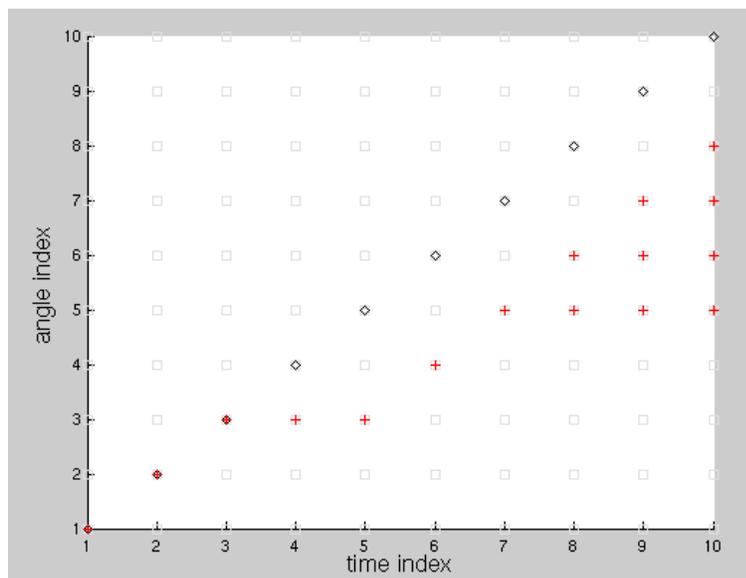}}
\caption{Possible trajectories through gantry angle/time space.}
\label{gridFig}
\end{figure}

The difficulty in solving the both-ways model stems from the large number of variables $b$.  The following modification
allows us to restrict the gantry trajectory as we see fit.  For example, if we would like the gantry to stay exactly two
time units at every even angle index, and three time units at every odd index, this would be possible.  
We imagine the gantry progression as plotted on a gantry position versus time plot, see Figure \ref{gridFig}.  
The squares represent the possibilities one needs to account for in the both-ways model, when any angle is a possible
starting angle. The diamonds, on the diagonal, represent the restricted model where the gantry starts at angle 1 and 
goes forward without slowing down ($a=t$).  The cross hatches refer to something in between.  In this case the trajectory
is partially specified (go forward until angle index 3, then stay there for a total of 3 time units, then go forward
to angle index 5, at which point you can go forward or stay still for each of the remaining time units).  To model this
type of specification in general, for each angle we define a list of possible times (or equivalently, for each time we define a list
of possible angles).  Also, depending on restrictions of the direction of gantry motion, and for each valid (time, angle) pair, 
we define a list of possible previous angles.  For the cross hatch pattern
we will have PosTimes[6] (possible times for angle 6) = 
\{8, 9, 10\}, PosAngles[9] =\{5, 6, 7\}, and PPA[9,6] (possible previous angles for grid point time 9 and angle 6) = \{5, 6\}, etc.
(Note specifying all of PosTimes, PosAngles, and PPA is redundant but helpful in the programming of this model).
With these sets defined, we can then rewrite constraint (\ref{validNextAp}) as 
\begin{equation}
b[l,r,a,t] \le \sum_{p \in \mathrm{PPA}[t,a]} ~~~ \sum_{i =\mathrm {max}(0,l-1)}^{l+1} ~~~ \sum_{j = \mathrm {max}(i,r-1)}^{\mathrm {min}(M,r+1)} b[i,j,p,t-1].
\label{validPPA}
\end{equation}
These constraints are written for all $l = 0\ldots M,~ r = l \ldots M,~ t = 2 \ldots T$ and $ a \in \mathrm{PosAngles}[t]$.

\subsection{Lower bound model (DAGU)}
Since we are unable to solve the both-ways model without additional restrictions, we instead 
solve the PPA model (we discuss below how we set the PPA data for the runs).  
For specific instances, we would like to know how far our PPA solution is
from the optimum of the both-ways model?  To address this, we  build a model which serves as
a lower bound to the both-ways model.   The simplest lower bound model is the base IMRT model given in formulation
(\ref{basemod}).  This is a lower bound because it assumes we have the time to do full intensity modulation at each angular position.
To tighten this lower bound, we will add some delivery information into the IMRT model.

The more jagged an intensity profile at a single angle is, the longer it will take to deliver it.  When leaf speed is not
considered, this idea can be included in linear optimization models as the sum-of-positive gradients (SPG) \cite{craft-spg}.  
In our model, we do consider leaf speed, but still, adding in an SPG constraint will move us in the right direction.
For each angle $a$ we define
\begin{equation}
\mathrm {SPG}[a] = x[\mathrm{start}(a)]+
\sum_{k=\mathrm{start}(a)+1}^{\mathrm{end}(a)}(x[k]-x[k-1])_{+} \\
\label{spg}
\end{equation}
\noindent where start$(a) = (a-1)M+1$ and end$(a) = aM$ are the starting and ending beamlet indices
for angle $a$, and the notation $(\cdot)_{+}$ is shorthand for max(0, $\cdot$).
Note that this sums up positive gradients, i.e. positive jumps in the beamlet intensities, 
across the beamlet row.  Although it does not give the delivery time,
it does provide an estimate of the delivery time that is always too low (since leaf speed is considered infinite) and that
we can make use of.

We also know that if two non-adjacent angles are both used but the angles between them are not used, then although
the SPG for those unused angles will be zero, it still takes one unit of time for the gantry to pass each one of them.  
To handle this
we use a binary variable called DAGU which stands for `does angle get used'.  We also use a variable called UTA which
stands for `up to angle', which gives the maximum angular position that the gantry used in a delivery.
Assuming DAGU and UTA are properly set, we can then include the constraint
\begin{equation}
\sum_{a=1}^A \mathrm {SPG}[a] + \mathrm {UTA} - \sum_{a=1}^A \mathrm{DAGU}[a] \le T.
\label{TCON}
\end{equation}
The rationale behind this inequality is that for every angle that gets swept through but not used (there are
UTA - $ \sum_{a=1}^A \mathrm{DAGU}[a]$ of them) we include one unit of time to account for the finite gantry speed.
For the angles that do get used, we use their SPG measure.
To set the DAGU variables, we use the following constraints.
\begin{equation}
\mathrm{DAGU}[a] \le \sum_{k = \mathrm{start}(a)}^{\mathrm{end}(a)} x[k], ~~~~~ a = 1 \ldots A
\end{equation}
If all the x values in a beamlet row are 0 then this forces DAGU to 0.  Also,
\begin{equation}
x[k] \le \mathrm{DAGU}[\mathrm{floor}((k-1)/M)+1]*T, ~~~~~ k = 1\ldots M*A,
\end{equation}
which forces beamlets to 0 if the DAGU for that beamlet is 0.  Otherwise if DAGU for that angle is 1
then the beamlets can take on arbitrarily high values (but any one beamlet value can never
be above the treatment time $T$).
The following inequality, in conjunction with inequality (\ref{TCON}), forces UTA to the correct value:
\begin{equation}
\mathrm{UTA} \ge a\cdot \mathrm{DAGU}[a], ~~~~~ a = 1\ldots A.
\end{equation}
Finally, we specify that $x$ needs to be integer valued.  
Although this formulation is also a mixed integer program (due to the integrality of $x$, DAGU, and UTA), 
there are far fewer integer variables making it much more tractable.

\subsection{Optimization procedure}
We have found through experimentation that the results of the lower bound (DAGU) runs are helpful
for specifying the input for the PPA problem.  Specifically, one of the outputs from a DAGU run is a list of
the SPG values at each gantry angle.  We use these SPG values directly by setting the time that the
gantry stays at that position equal to the SPG value.  For angles with SPG equal to 0, but subsequent angles with
positive SPG, we set the gantry to stay there
for exactly one time step.  We stop the beam at the highest angle used by the DAGU run.  Thus we specify
the gantry motion dynamics exactly from the DAGU analyses output.  It is not clear a priori that this is 
the best strategy, although it does make intuitive sense to slow down the beam in proportion to the complexity 
of the intensity map delivered there.  We have found that this method is better than
other alternatives, including allocating the time approximately equally across angles, or giving more time
to angles with SPG $>1$ but not considering the actual value of the SPG beyond that.
The initial gantry position is set to  0 degrees -- the 
posterior/anterior beam angle -- for all VMAT runs.
\section{Results}
\begin{figure}[ht] 
\centerline{\includegraphics[width=10cm]{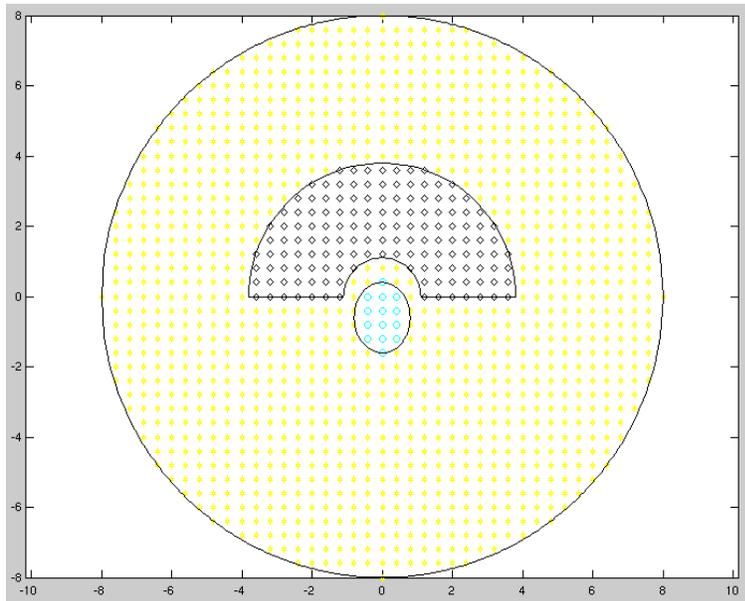}}
\caption{Phantom geometry used.  The horseshoe structure is the target and the elliptical set of voxels below represents
the critical structure.  Voxels are 0.4 cm by 0.4 cm, and the scale of axes is cm.}
\label{geomFig}
\end{figure}

We run the main set of optimizations on the phantom geometry shown in Figure \ref{geomFig}.
For the dose optimization, we consider a bi-criteria problem where the two objectives
are 1) minimize the maximum dose of the target 2) minimize the mean dose of the critical structure, which we 
will call the cord.  Our objective function is then $f = w_t (\mathrm {tumor~max}) + w_c (\mathrm{cord~mean})$,
where $w_t$ and $w_c$ are objective function weights.
We constrain the target voxels to receive at least 50 units of radiation, 
and all voxels no more than 100 units (these values were chosen based on the target dose profile for an open field
radiation delivery of full gantry rotation around the phantom at maximum speed).

We use $M=4$ beamlets on the gantry head, and $\Delta = 4$ degrees, i.e. 90 angles around the patient,
beamlet size $s = 2$ cm.  To assess
the tradeoffs of dose distribution quality and treatment times, we run every combination of the following possibilities:
treatment time $T$ within the set \{90, 135, 180\} (these times represent the time needed to rotate the gantry 
once around the patient, one and a half times, and twice, twice being equivalent to allowing the gantry to stop for two time
units at each angle),  and the objective function weighting vectors $(w_t, ~w_c) = (1,~0), ~(0.5,~0.5),$ 
and $(.1, ~.9)$.  This gives a total of 9 combinations, and for each case we run both the lower bound (DAGU) and the upper 
bound (PPA) models.  Additionally, for each of the weighting vectors, we run the IMRT optimization (every angle, every beamlet, 
no constraints on the $x$ vector except for non-negativity) as an ideal (i.e. unrestricted delivery time) lower bound solution.  
The final weighting vector of $(.1, ~.9)$ was chosen instead of $(0, ~1)$ since $(0, ~1)$ leads to highly unrealistic 
plans regarding tumor hot spots.


\begin{figure}[ht] 
\centerline{\includegraphics[width=10cm]{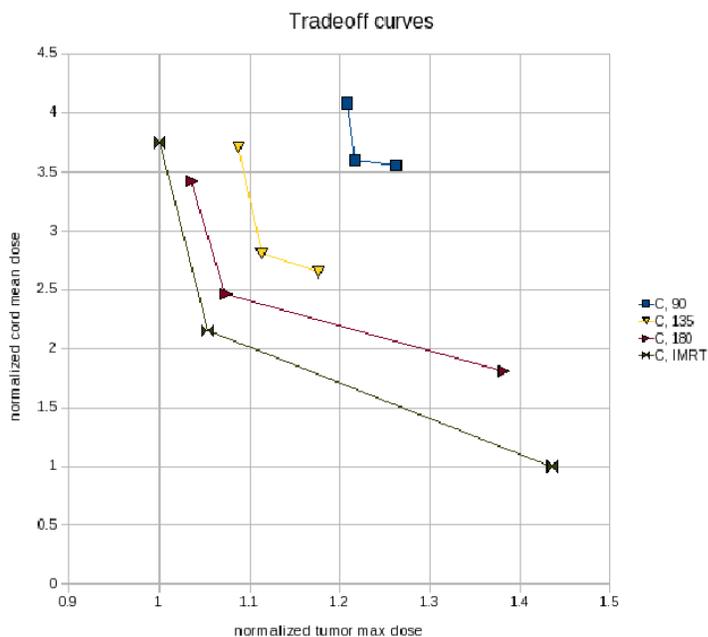}}
\caption{Dosimetric tradeoffs (tumor maximum dose versus cord mean dose) for different time settings, for the best deliverable
plans found.  The IMRT solution
represents the ideal case where time is not considered and full intensity modulation can be done at each gantry angle used.
For this analysis, 90 time units is the time for the gantry to travel once around the patient at maximal speed, i.e.
only spending one time unit at each angle modeled.  }
\label{trades}
\end{figure}

In Figure \ref{trades} we plot the dosimetric tradeoff curves (these are the results of the PPA runs, i.e. 
the best deliverable plans found) for different treatment times, as well as the ideal
tradeoff if full intensity modulation is allowed at every discrete angle modeled.  
We see that for the $T=180$ case the maximal tumor dose is reduced to within 4\% of its ideal IMRT value,
but it is harder to get the cord mean dose down (it is 80\% higher than its ideal IMRT value) with this 
amount of time, due to the highly concave dose distribution needed in that case.  As the treatment time is further reduced,
both objective functions are worsened increasingly, particulary for the cord.

\begin{figure}[ht] 
\centerline{\includegraphics[width=17cm]{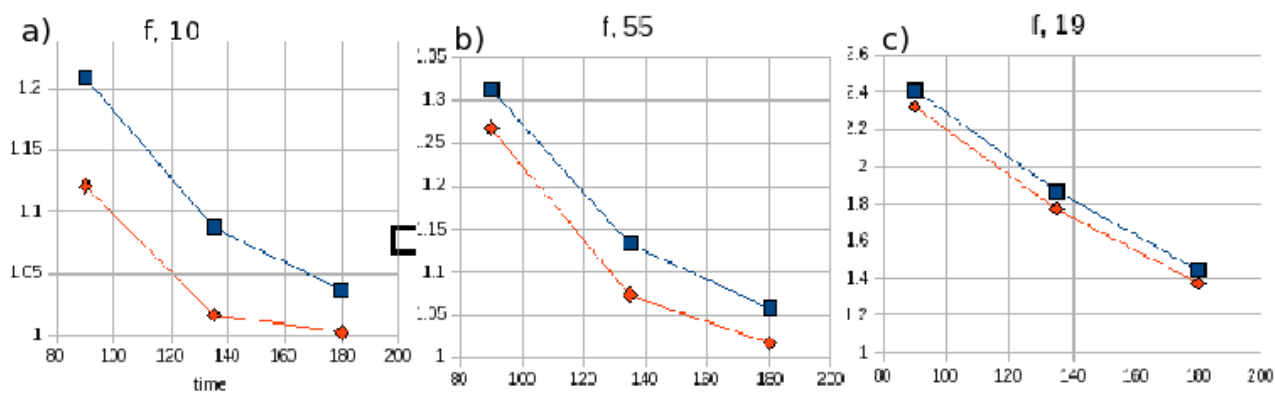}}
\caption{Tradeoff curves of optimization function value versus time for the three weight vectors used.  
a) is for  $(w_t, ~w_c) = (1,~0)$, b) is (.5, .5), and c) is (.1, .9).  Thus, as we move to the right, the 
optimal dose distribution becomes more concave, requiring more intensity modulation. Upper curves
are the best feasible VMAT solutions found, using the PPA method described in the text.  Lower curves are 
the lower bound from the DAGU method.}
\label{105519}
\end{figure}

\begin{figure}[ht] 
\centerline{\includegraphics[width=14cm]{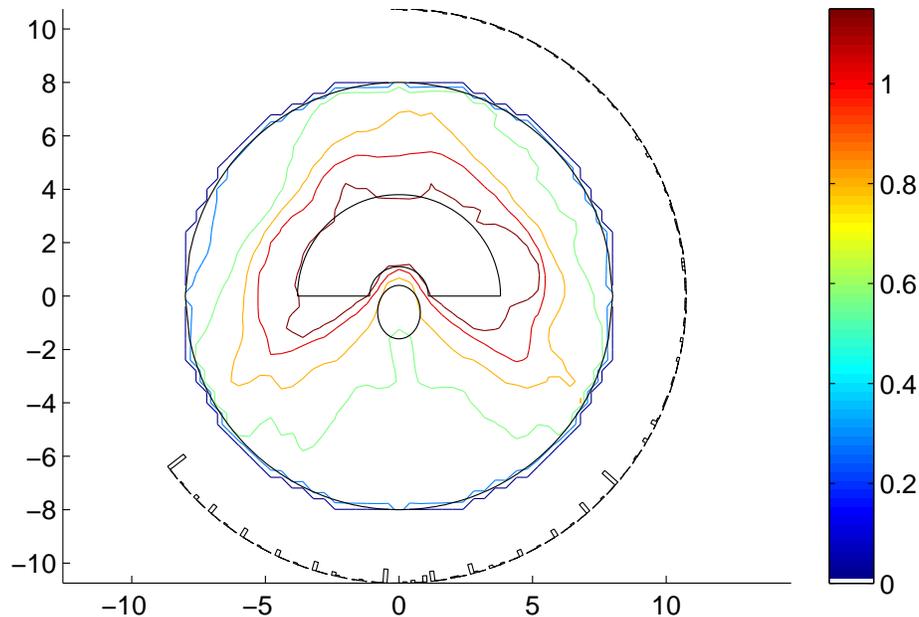}} 
\caption{Dose distribution and intensity profiles at each beam direction for $(w_t, ~w_c) = (0.5,~0.5)$ and the time $T=180$.
The dose has been normalized by the prescription dose of 50.  At each angle we miniaturize the beam head and plot the 
intensity of the 4 beamlets, so that the modulation can be shown without overlapping.}
\label{doseAndRing}
\end{figure}

We can also view this information by plotting, for each fixed objective weighting vector, the tradeoff between plan quality, 
as measured by the objective function, and treatment time.  Figure \ref{105519} shows this for the three objective weight 
vectors used, for both the upper and lower bounds.  It is clear that extra time is much more beneficial when a non-convex
dose distribution is required, i.e. when minimizing the mean cord dose is a significant component of the objective function.
Also note that in all cases, the upper and lower bounds are within 10\% of each other, and for the two cases where the
cord weight is positive (the more realistic cases), the bounds are within 7\% in all cases.  It is also clear from Figure
\ref{105519}(c) that more time would be beneficial for this highly concave ideal dose distribution since the solution given
for the largest time $T=180$ is still over 40\% worse than the ideal IMRT solution.  
For the case  $(w_t, ~w_c) = (0.5,~0.5)$ and the time $T=180$, 
we show the deliverable solution (i.e. the upper bound solution provided by running the PPA model) in 
Figure \ref{doseAndRing}.  At each angle we plot the fluence map delivered at that angle, scaled down geometrically
so there is no overlap of the beamlets (length of rectangles are used to indicate the relative 
value of the fluences).  
One sees that the optimal solution for this case involves only a small amount of
radiation delivered in the first 1/4 revolution, and for the next 135 degrees, highly intensity modulated beams are 
delivered every 8 degrees or so, with the MLC leaves blocking the cord.

Regarding the use of the DAGU results to set up each PPA run, we note that, for example, for the case
$(w_t, ~w_c) = (0.5,~0.5)$ and the time $T=180$, if we instead force the gantry to spend exactly 2 time units
at each angle, the resulting optimal solution is 5.5\% higher.  Similar values were found for the other cases.

\begin{figure}[ht] 
\centerline{\includegraphics[trim=0mm 0mm 0mm 0mm, clip,width=15cm]{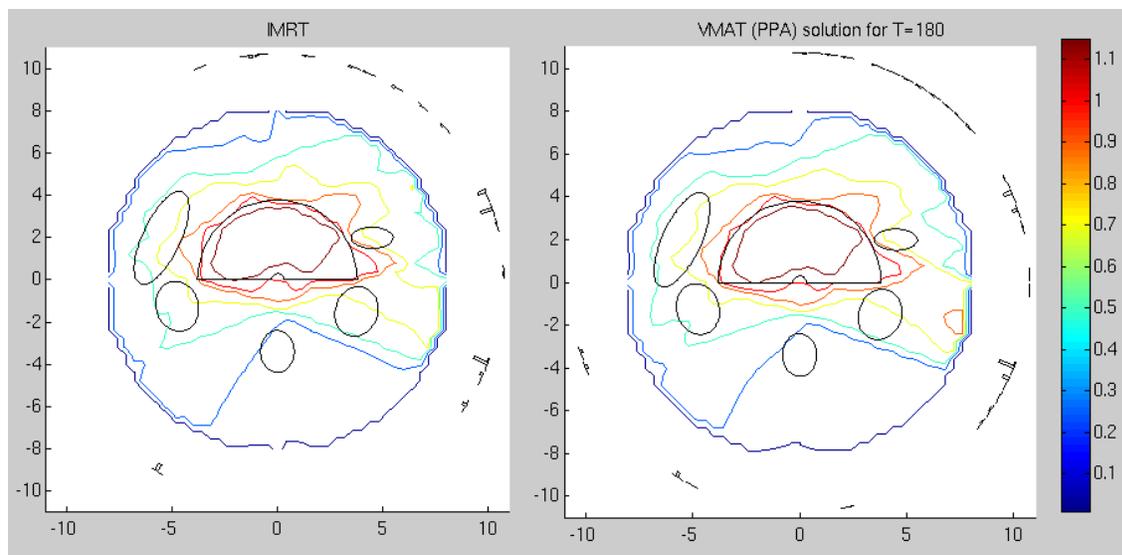}}
\caption{Results for the pancreas phantom.  The left plot shows the ideal IMRT solution and the optimal fluences.
On the right shows a deliverable VMAT solution obtained with the PPA model, for $T=180$.  The dose plots are almost
identical since this is enough time to deliver the required fluence pattern with VMAT.  The fluence maps
are similar also.}
\label{pancCase}
\end{figure}

We also apply the methods to a phantom case which has more structures.  This case mimics a pancreas case, and is shown
in Figure \ref{pancCase}.  
The objective function is the sum of the target maximum dose plus the mean doses from each critical structure.
The deliverable solution shown on the right of the plot is for $T=180$.
It is apparent from the similarity of this dose distribution to the IMRT dose distribution (the objective 
functions differ by 0.3\%) that this is enough time
to deliver a near ideal solution.  Also, the entry angles for the doses are nearly the same between the IMRT plan 
and the VMAT plan.  The DAGU run (not shown, but used to initialize the PPA run) also had a nearly identical fluence map.
When the time was reduced to $T=90$ (also not shown), the DAGU solutions and PPA solutions were again almost indistinguishable, although
they were both farther from the IMRT solution (5\% worse).

\section{Discussion and conclusions}
It is common in radiotherapy optimization studies to ignore radiation delivery time.  This is partially due
to the two-step optimization method often used in IMRT planning.  In step one, optimal beamlet fluences are 
computed.  In step two, those fluence maps are processed into MLC leaf sequences.  Since the dosimetric plan 
quality is handled in step one, and the delivery time is given by the results of step two, it is not common
to view the problem coupled.  However, when VMAT promises to deliver quality radiation therapy plans in short
times, it will be prudent to include the timing information in plan comparisons, so physicians and hospitals 
can make the proper delivery choices.

One solution would be to have the treatment time $T$ included directly as constraint in the treatment plan 
optimization, and let the optimizer choose the best delivery method, subject to machine constraints, given the
dosimetric problem formulation.  To solve such a problem to proveable optimality is very difficult, and likely
beyond algorithmic and computational resources for years to come.  A good alternative then is to at least provide
the delivery time when doing plan comparisons.   

In this paper, using a simplified model -- a 2D phantom and a spatially and temporally coarse MLC/gantry model -- 
we have included delivery time as a constraint and we have not specified to the optimizer whether to do IMRT or VMAT.
Instead, the optimizer chooses.  When the optimal isodose curves are non-convex, treatment time becomes critical as it takes
more intensity modulation from specific angles to produce the right dose distribution.  In these cases (the first set of runs, 
with a single critical structure having positive objective function weight), 
we have found that when the beam is moving from one angle to the next,
it does not fully close the MLC. In the pancreas phantom case however, the optimal dose distribution is more convex, and
time is less important and the MLC stays closed for much of the arc around the patient. 
Thus, IMRT with suitable angles is prefered if time is not an issue.  When delivery time is a consideration
it is wasteful to move the beam while not delivering any dose, and VMAT will be the preferred delivery technique.  
But to reiterate, in an ideal planning world, this decision would be made by the software, or the software would provide optimal
solutions for various values of treatment time, allowing the physician to select tradeoff between plan quality and treatment
time.  Also, if the gantry is allowed to stop at any angle for an unrestricted time, and the beam can be fully blocked
when the gantry is rotating, VMAT is a superset of IMRT.   In that case, the discussion between modalities is irrelevant.
In practice though, one needs to make the decision between IMRT and VMAT, and the results of this paper suggest that
VMAT should be used when minimizing delivery time is paramount;
otherwise, IMRT from well-chosen angles is the best choice, and solving the IMRT problem with many possible angles
may yield a small subset of collectively promising angles.

From the planning perspective, a significant 
benefit of VMAT is that it does not require beam angle optimization.  However, to maximally benefit from VMAT, 
the optimization of VMAT plans needs to be strong, and ideally provide an estimate of how close to the true optimum the 
plan is.  The lower bound provided in this paper would be good for this purpose except for the fact that
for clinical sized problems, this model will not likely be solvable.  A convex alternative to the lower bound formulation
could involve using the SPG at each angle as a proxy for the delivery time there, and dropping the integrality on $x$.
Following the method demonstrated in this paper -- including an SPG constraint in the IMRT model -- 
one can get an idea of where one should keep the gantry for longer periods.
This input may be very useful in VMAT optimization.

\medskip
\medskip
\medskip
\medskip
\medskip
\bibliographystyle{plain}
\bibliography{all}

\end{document}